
\magnification=\magstep1
\hsize=15truecm
\vsize=22truecm
\hoffset=0.5truecm
\def\double{\baselineskip=0.9truecm}

\raggedbottom
\font\text=cmr10
\def\ltsim{\mathrel{\hbox{\rlap{\hbox{\lower4pt\hbox{$\sim$}}}\hbox{$<$}}}}
\def\gtsim{\mathrel{\hbox{\rlap{\hbox{\lower4pt\hbox{$\sim$}}}\hbox{$>$}}}}

\text

\def\apj    {{\it Ap.~J.~}}
\def\apjl   {{\it Ap.~J.~(Letters)~}}

\def\aj     {{\it A.~J.~}}
\def\aa     {{\it Astr.~Ap.~}}
\def\mnras  {{\it M.N.R.A.S.~}}

\def\pasp   {{\it P.A.S.P.~}}

\def\ref {\noindent}
\def\v#1  {{\bf {#1}~}}

\def\etal {{\it et al.~}}

\double

\centerline {\bf Red Giants in the Halo of the S0 Galaxy NGC 3115: }
\centerline {\bf A Distance and a Bimodal Metallicity Distribution$^1$} 

\vskip 2cm

\centerline {Rebecca A. W. Elson }
\centerline {\it Institute of Astronomy, Madingley Road, Cambridge CB3 0HA,
 England }
\centerline {elson@ast.cam.ac.uk}
\vskip 6cm

\noindent
Submitted to MNRAS February 29, 1996

\noindent
Revised October 4, 1996; Accepted Novemvber 25 1996

\vskip 4cm

\noindent
$^1$ Based on observations obtained as part of the Medium Deep Survey.

\vfill\eject
\centerline {\bf Abstract}

Using the Hubble Space Telescope, we resolve the red giant branch in the
halo of the S0 galaxy NGC 3115.  We measure magnitudes and $(V-I)$
colours for stars down to 1.5 magnitudes below the tip of the red giant
branch.  From the brightest stars we estimate a distance modulus 
$(m-M)_0=30.21 \pm 0.30$, corresponding to a distance of $11.0 \pm 1.5$ 
Mpc.  This is
in excellent agreement with the value $(m-M)_0=30.17 \pm 0.13$
determined from the planetary
nebula luminosity function.  Our results rule out the shorter distance
modulus $(m-M)_0=29.65$ determined from surface brightness fluctuations.
A histogram of $(V-I)$ colours shows a clear bimodality, indicating the 
presence of two distinct halo populations of roughly equal size.  One 
has [Fe/H]$\sim -0.7$ and one has [Fe/H]$\sim -1.3$.  This is the most
distant galaxy in which a Population II halo has been resolved, and it
is the first
time a colour bimodality has been observed among the halo stars of 
any early-type
galaxy.  


\vskip 1cm
\noindent {\bf 1. Introduction}
\smallskip

With the advent of the Hubble Space Telescope (HST), the brightest giants in
galaxies as far away as the Virgo cluster are being resolved.  Since
the magnitude of the tip of a Population II 
red giant branch (the TRGB) 
is relatively insensitive to metallicity or age, 
it therefore 
provides a reliable distance indicator (cf. Lee, Freedman \& Madore 1993). 
This method 
has been applied to a variety of galaxy types within 
$\sim 1.5$ Mpc using ground based observations, (Lee \etal 1993; Sakai,
Madore, \& Freedman 1996), 
and more
recently, using HST, to NGC 5128 at 3.6 Mpc (Soria \etal 1996).
A good discussion of the method for determining
the magnitude of the TRGB, and the uncertainties introduced by photometric
errors, crowding, population size, and background contamination is given by 
Madore \& Freedman (1995).
 
As the accuracy of the photometry of such distant stars improves, we should
obtain, in addition to better distance estimates, a more thorough
understanding of 
the nature of stellar populations in the halos
of early type galaxies, of which we have no local examples.  For example,
the colour of the red giant branch should indicate the range of metallicities
present. Soria \etal find that the colour of the red giant branch
in the halo of 
NGC 5128 indicates a broad range of metallicities, with a mean value 
[Fe/H]$ > -0.9$.  Until now, such metallicity estimates have relied only on the 
interpretation of integrated light.

In this paper we  present photometry for $\sim 4700$ halo stars in
NGC 3115, an 
edge-on S0 field galaxy 
thought to be an
extinct quasar with a $\sim 10^9 M_\odot$ 
black hole at its centre (cf. Kormendy \etal 1996).  We use the magnitude
of the TRGB to estimate the distance, and the distribution of colours
to explore the chemical compostion of the halo. 
The data, obtained with HST as part of the
Medium Deep Survey (cf. Griffiths \etal 1994),
are described in Section 2. Determination of the distance to NGC 3115
is discussed in Section 3, and the metallicity of the sample
is analysed in Section 4.  The results are summarized in Section 5.

\vskip 1cm
\noindent
{\bf 2.  The Data, Sample Selection, and Photometry }
\smallskip

This study is based on a field 8.5 arcmin ($\sim 30$ kpc) 
east of the centre of NGC 3115, and 5 arcmin ($\sim 18 $ kpc) away from the
major axis of the (edge-on) disk.
Exposures in two filters (F814W$= I_{814}$ and F606W$=V_{606}$) were acquired on
1994 December 27.  A total of 11 exposures were obtained in $I_{814}$,
with a total exposure time of 6.4 hours, and 
10 in $V_{606}$, with total time 4.6 hours.
The images in each passband were co-added with a median
filter.  
Only the three Wide Field Camera chips were considered.
They have image scale 0.10 arcsec pixel$^{-1}$, 
and total area, excluding borders,
of 4.56 arcmin$^2$.  The FWHM is $\sim 1.5$ pixels. 
In order to ensure consistent selection criteria, we did not use the PC
chip;  with area only 0.36 arcmin$^2$ it would increase our sample
size by less than 10\%.

We 
used the DAOPHOT task DAOFIND to detect automatically all stellar
objects in the deeper $ I_{814}$ image.  We set the detection threshold to 
$3\sigma$.  In practice, DAOFIND detects, in addition to stars, a large
fraction of the background galaxies in the field, as well as spurious
sources along diffraction spikes and structure in the point-spread function
(PSF) surrounding bright (foreground) stars.

To discard non-stellar detections, we  constructed a PSF from
two bright, unsaturated foreground stars in chip 2, and fit this to all
objects detected in the $ I_{814}$ image, 
using a fitting radius of 10 pixels.  
We then used the output fitting parameters from ALLSTAR to isolate a sample of
stars, and eliminate most background galaxies and spurious detections.  
Figures 1a and b show the ``sharpness" and the Poisson error in
the magnitude 
plotted against $ I_{814}$ for all detections. 
The sharpness parameter measures the ratio of heights of
the best fitting delta function and Gaussian function.
Also shown are artificial stars created and measured using the same 
procedure as for the real objects.  

Figure 1a shows a concentration of objects around sharpness=0, which
are consistent with being stars.  A secondary sequence is also visible around
sharpness $\sim$0.1.  In Figure 1b there is  a concentration of 
points with larger errors at a given magnitude
than expected on the basis of
simulations.
Inspection of these outlying points reveals that they are either
spurious detections or background galaxies.
(A very bright foreground star just out of the field is responsible for 
pronounced diffraction spikes running across two of the chips.)
Our selection criterion for stellar objects is illustrated with
the solid lines in Figs. 1a and b.  
Our final photometry list contains 
$\sim 4700$ candidate
giant branch stars.  

It is our experience that with stellar photometry on
poorly sampled images, better results (as judged by the narrowness of
the stellar sequence in a colour-magnitude diagram (CMD)) 
are often achieved with aperture
photometry rather than with PSF fitting.  Experiments showed that this 
is indeed the case here.  Our final magnitudes were therefore measured
with an aperture of radius 2 pixels.  We applied the following corrections:
an absorption $A_I=0.05$ (Ciardullo, Jacoby \& Tonry 1993); an adjustment for 
geometric distortion; an aperture correction to transform from 2 pixel
to total (r=10 pixel) magnitudes. Finally, we converted the
instrumental `flight system' magnitudes to the Johnson-Cousins system.
All the corrections and transformations were applied following 
Holtzman \etal (1995a,b). 

Figure 2 shows a CMD 
for the $\sim 4700$  objects.  There is a clear cutoff 
at $I_0 \sim 26$, and only a few objects brighter than this. The majority
of these are very blue, and are probably compact background galaxies
(cf. Elson, Santiago \& Gilmore 1996).  
A few may be blends of two individual giants, and a few with 
$(V-I)_0 \gtsim 2$ may be AGB stars.

\vskip 1cm
\noindent
{\bf 3. Determining the Distance to NGC 3115}

As long as the stars in our field are metal poor ([Fe/H]$\ltsim -0.7$),
we may assume an absolute
magnitude of $M_I=4.00 \pm 0.10$ for the TRGB, 
regardless of the precise metallicity of the population (Lee \etal 1993).
Since our field lies well away from the NGC 3115 disk, it should be
dominated by halo stars.  Colour gradients measured along the minor axis
of NGC 3115 show that the $(B-V)$ colour becomes bluer with
radius, and by $r\sim 5$ kpc, $(B-V) \sim 0.8$ (Strom \etal 1976).  
This is only 0.1 mag
redder than the mean for Galactic globular clusters, which have mean
metallicity well below [Fe/H]$=-0.7$.  The assumption that [Fe/H]$\ltsim
-0.7$ in our halo field is further supported by surface photometry in 
$B$ and $R$ by Surma, Seifert \& Bender (1990), and by our own results
in Section 4 below.  

Our primary task is therefore to determine the precise magnitude of the
TRGB, and to estimate its uncertainty.  We adopted as the TRGB the
brightest 
magnitude at which the giant branch 
luminosity function steepens dramatically.
To quantify this, we calculated a derivative across bins of 0.1 to
0.4 mag for luminosity functions drawn from different samples.  We also
used a Sobel filter 
as described
in Madore \& Freedman (1995), with software kindly provided by S. Sakai. 
All the methods gave values converging on $I_0 = 26.15$, as we now 
discuss.

Figures 3a-c show the value of the derivative of the luminosity function
plotted against $I_0$ for the following samples:

(a) The full sample with steps $0.1-0.4$ mag.

(b) Subsamples from chips 2, 3 and 4 with step 0.2 mag.

(c) The sample of objects redwards of the dashed line in Fig. 2, with
steps 0.1, 0.2 and 0.3 mag.

\noindent
Figure 3d shows the output from a Sobel filter, again for the three
individual chips, for the red sample, and for the full sample.

Figures 3a-c show two, or in some cases three prominent 
peaks.  In the cases where two peaks are present, their values, 
averaging over all cases, are $I_0=26.14 \pm 0.04$ and $I_0=26.41 \pm 0.06$.
In a fewe cases there is a brighter, smaller, peak with $I_0=25.92 \pm 0.02$.
The first main steepening in the luminosity function thus occurs at 
$I_0=26.14\pm0.04$.
If instead we adopt the mean value of the three peaks, we get
$I_0=26.16 \pm 0.25$.  The values do not differ significantly, and we
conservatively adopt the latter, with its larger error.

Adjustments to the measured magnitude of the TRGB may be 
required due to noise, crowding, small numbers,
and background contamination.  Madore \& Freedman assess these systematic
uncertainties, and we rely on their results to determine any 
necessary corrections.  In our sample,
the typical signal-to-noise ratio in stars with $I_0=27$ 
is $\sim 35$, which is sufficient, according to Fig. 3 of Madore \& Freedman,
not to introduce any systematic error in the determination of the 
magnitude of the TRGB.

We assessed the effects of crowding by 
populating a blank frame with artificial stars according to 
a luminosity function of the form 
$\log N = 0.6 I$ (cf. Madore \& Freedman 1995), 
normalized to the same surface density of stars in
the mangitude range $26 < I_0 < 27$ as observed in our field.
We considered ``blends" to be stars within 2 pixels of each other.
Only 1\% of stars in the magnitude interval $26 < I_0 < 27$
were blended with other stars in the same interval.  14\% of stars
in this interval were blended with stars up to two magnitudes 
fainter.  
This is consistent with what we would estimate based on
the surface 
brightness of NGC 3115 at the position of our field 
and the effective seeing of HST.
Referring to Fig. 6 of Madore \& Freedman, this 
suggests that the apparent magnitude of the TRGB is brighter than the
true value by $\sim 0.05 \pm 0.05$ mag. 

Our CMD is sufficiently well populated that we would
not expect population size to introduce any uncertainty into our determination
of the magnitude of the TRGB. 
There are $\sim 2700$ stars with $26.15 < I_0 < 27.15$ in Fig. 2.  
Figure 7
of Madore \& Freedman implies that 
any systematic effects due to population size
should be negligible. 

Contamination from foreground stars in our own galaxy, and from globular
clusters or Population I stars in NGC 3115 is negligible:  both foreground
stars and globular clusters are rare; only a few are expected in the
entire field, and these will be much brighter than the TRGB.
The main source of contamination at these magnitudes is compact
galaxies.
We know from the Hubble Deep Field that unresolved 
galaxies are predominantly blue, with $(V-I)_0 < 1.0$ (cf. Elson, Santiago
\& Gilmore 1996).
Applying the derivatives and edge-detection algorithm only to 
objects redder than the dashed line in Fig. 2 gives Fig. 3c and the
dotted curve in Fig. 3d.  Background contamination is evidently not 
important in our determination of the magnitude of the TRGB.

We also checked the effect of slightly different selection lines in
Figure 1a.  This contributed an uncertainty of $\pm 0.05$. 
Thus, the main error is the photometric uncertainty of $\pm 0.12$ at 
$I_0=26.16$, and the uncertainty in isolating the point of steepest
slope of the luminosity function, of  $\pm 0.25$.
Combining these, and correcting our best value for crowding as above,
the final value we adopt for the TRGB is $I_0=26.21 \pm 0.29$.
Including the error of $\pm 0.10$ in the absolute magnitude of the TRGB, 
implies a distance modulus of $(m-M)_0=30.21 \pm 0.30$, corresponding
to a distance of 11.0$\pm 1.5$ Mpc.  This is in excellent agreement
with the value $30.17 \pm 0.13$
quoted by Ciardullo, Jacoby \& Tonry (1993) from the planetary nebulae
luminosity function. It is 0.56 mag greater than the value $29.65 \pm
0.25$ they quote derived 
from surface brightness fluctuations.  The latter would
imply a magnitude for the TRGB of $I_0=25.65$, which is inconsistent
with our observations, as is apparent from an inspection of Fig. 2. We
suggest, therefore, that this value may be wrong.   We note
however that in the sample of 16 galaxies reported in Ciardullo \etal
(1993), the surface brightness fluctuation 
value for NGC 3115 was the most discrepant one,
and is not indicative of the general accuracy of the method.

\vskip 1cm
\noindent
{\bf 4. The Metallicity of the NGC 3115 Halo}

The colour of giant branch stars is a function of their metallicity. 
This is illustrated in Fig. 2, where the loci of giant branches
in three Galactic globular clusters (Da Costa \& Armandroff 1990)
are transposed to a distance modulus of 30.2
and superposed on the NGC 3115 data.
The globular clusters have a wide range of metallicity:  the one with 
the bluest giant branch, M15, has [Fe/H]=$-2.17$, the intermediate one,
NGC 1851, has [Fe/H]=$-1.29$, and the reddest one, 47 Tuc, has 
[Fe/H]$=-0.71$.  At a glance, it is clear that the halo of NGC 3115 is not
very metal poor, and that it has [Fe/H]$\sim -1$.

To determine the value more precisely, we constructed $(V-I)$ histograms in 0.2
mag intervals.  These are shown in Figs. 4a-d.  Here we encountered the 
surprising result that {\it the colour distribution is bimodal}.  This is
particularly evident in (c) where the sample is biggest.  We fit two gaussians
in the range $1.0 < (V-I)_0 < 2.5$ allowing the amplitude, mean colour, 
and sigma
to vary.  The best values of $<(V-I)_0>$ in each case are plotted as large
dots in Fig. 2.  The bluer population is well represented by the 
giant branch with [Fe/H]$\sim -1.3$.  The redder population appears to have
[Fe/H]$\sim -0.7$.  The ratio of the sizes of the blue to red
populations is 1.3, 2.4, and 3.4 in the faint, intermediate, and bright
bins respectively.  These ratios are of course uncertain, particularly
for the brighter samples, and we conclude only that the populations
are of equal size to within a factor of about two.  In all magnitude bins 
the value of $\sigma$ for the redder population is comparable to the
measurement errors, while that for the bluer population is somewhat
larger than the errors.  This may indicate an intrinsic spread in 
metallicity among the bluer population, or may be due to contamination
by unresolved background sources.  Our data neither contain evidence for,
nor rule out, 
different spatial distributions for the two populations. 

It is unlikely that the observed colour bimodality is due to the presence of 
populations with different ages.  In this case, if the red population
were old and metal rich (like 47 Tuc), then a younger population with
the same metallicity would have to 
be only $\sim 1$ Gyr old (and therefore much brighter).  The presence of such a 
young population in the outer halo of an S0 galaxy would be difficult to 
explain. 

\vskip 1cm
\noindent
{\bf 5. Summary}

We have presented  a CMD for $\sim 4700$ 
giant branch stars in the halo of the S0
galaxy NGC 3115.  The magnitude of the TRGB indicates a distance modulus
of $(m-M)=30.21 \pm 0.30$, in excellent agreement with the value 
$30.17 \pm 0.13 $, derived
from the planetary nebula luminosity function.  Our observations
are not consistent with the smaller distance modulus $29.65 \pm 0.25$
derived 
from surface brightness fluctuations.

The $(V-I)_0$ colour distribution of stars near the TRGB in the NGC 3115
halo is bimodal.  We interpret this as indicating the presence of two
distinct populations, one with [Fe/H]$\sim -0.7$ and one with 
[Fe/H]$\sim -1.3$.  In comparison, the halo of the Milky Way has 
[Fe/H]$\sim -1.7$ (Ryan \& Norris 1993), while the halo of M31 
appears to be
more metal rich, with [Fe/H]$\gtsim -1$ (Mould \& Kristian 1993;
van den Bergh 1991). 
This is the first time such a bimodality has been 
observed in the stellar population of the halo of an early type galaxy,
although in some cases such as M87 
globular cluster systems exhibit such bimodality
(cf. Ajhar, Blakeslee \& Tonry 1994; Elson \& Santiago 1996).  
Our result suggests either that NGC 3115 is the product of a merger
between two galaxies of similar size, or that its halo underwent two
distinct episodes of star formation, the second of which was from 
gas enriched by a factor of $\sim 4$ compared to the first episode.

It would be interesting to see whether the globular cluster system of
NGC 3115 reveals a corresponding colour bimodality, and also whether,
with more accurate photometry, NGC 5128, another peculiar S0 galaxy,
would exhibit the same effect.

\bigskip
\noindent
{\bf Acknowledgements}

I would like to thank Barry Madore and Basilio Santiago 
for helpful discussions, and Shoko Sakai for making her software available.

\vfill\eject
\noindent{\bf References}

\ref
Ajhar, E., Blakeslee, J. \& Tonry, J. 1994 \aj 108, 2087

\ref
Capaccioli, M., Cappellaro, E., Held, E., \& Vietri, M., 1993 \aa 274, 69

\ref
Ciardullo, R., Jacoby, and Tonry, J., 1993 \apj 419, 479

\ref 
Da Costa, G. \& Armandroff, T. 1990 \aj 100, 162

\ref
Elson, R. \&  Santiago, B. 1996 \mnras 280, 971

\ref
Elson, R.,  Santiago, B., \& Gilmore, G., 1996 {\it New Astronomy} {\bf 1}, 1

\ref
Griffiths, R., \etal 1994 \apj 437, 67

\ref
Holtzman, J. A., \etal 1995a  \pasp 107, 156

\ref
Holtzman, J. A., \etal 1995b  \pasp 107, 1065

\ref
Kormendy, J. R. \etal 1996 \apjl 459, L57

\ref
Lee, M. G., Freedman, W., \& Madore, B. F., 1993 \apj 417, 553

\ref
Madore, B. F. \& Freedman, W., 1995 \aj 109, 1645

\ref
Mould, J. \& Kristian, J. 1986 \apj 305, 591

\ref
Ryan, S. \& Norris, J. 1993 ASP 48, {\it The Globular Cluster -- Galaxy 
Connection} 

eds. G. Smith \& J. Brodie (San Frransisco) p. 338

\ref
Sakai, S., Madore, B. \& Freedman, W. 1996 \apj 461, 713

\ref
Soria, R. \etal 1996 \apj 465, 79

\ref
Strom, K., Strom, S., Jensen, E., Moller, J., Thompson, L., \& Thuan, T., 1977

\apj 212, 335

\ref
Surma, P., Seifert, W. \& Bender, R. 1990 \aa 238, 67

\ref
van den Bergh, S. 1991 \pasp 103, 1053

\vfill\eject
\noindent
{\bf Figure Captions}

\vskip 0.5cm
\noindent
{\bf Figures 1a and b.}  
The parameters (a) ``sharpness" and (b) 
the magnitude error output by ALLSTAR, plotted against $ I_{814}$ magnitude
for $\sim 6500$ detections in the NGC 3115 field.  Crosses are artifical
stars.   Solid lines illustrate our selection criteria for stellar 
candidates.  Magnitudes are from PSF fitting with a 10 pixel fitting
radius. 

\vskip 0.5cm
\noindent
{\bf Figure 2.}
A colour-magnitude diagram for $\sim 4700$ 
objects selected using the
criteria illustrated in Figs. 1a and b. 
Magnitudes are from aperture photometry, dereddened and transformed to the
Johnson-Cousins system.
Most objects with $I_0 < 26$
are probably compact background galaxies.  Representative
Poisson error bars are shown. The two horizontal lines indicate the 
magnitude of the TRGB implied by the surface brightness fluctuation
method and the planetary nebula luminosity function method.
The curves are loci of giant branches of Galactic globular
clusters transposed to a distance modulus of 30.2.  They are for M15
([Fe/H]$=-2.17$), NGC 1851 ([Fe/H]$=-1.29$), and 47 Tuc ([Fe/H]$=-0.71$).
Large dots indicate the peaks of the histograms in Figs. 4a-c. 

\vskip 0.5cm
\noindent
{\bf Figures 3a-d.} Local slope as a function of $I_0$
of the NGC 3115 giant branch 
luminosity function for (a) the full sample with step sizes 0.1, 0.2, 0.3
and 0.4 mag; (b) the individual WFC chips with step size 0.2 mag;
(c) the subsample redder than the dashed line in Fig. 2, with step sizes
0.1, 0.2 and 0.3 mag.  (d) Output of the Sobel filter edge-detection algorithm
plotted against $ I_0$.  The solid curve peaking at $\sim 800$ is for the
full sample, and the dashed curve is for the redder sample. 
The curves peaking at $<400$ are for the individual WFC chips.

\vskip 0.5cm
\noindent
{\bf Figures 4a-d.} Histograms of $(V-I)_0$ for three different magnitude
bins.  Total number of stars are (a) 316, (b) 460, and (c) 639.
The vertical dashed lines indicate the peak values from 
Gaussian functions fitted in the range $1.0 < (V-I)_0 < 2.5$.  (d) Shows
the best fitting Gaussian functions for the histogram in (c).  Values of
$\sigma$ are 0.26 and 0.21 for the blue and red Gaussians respectively.

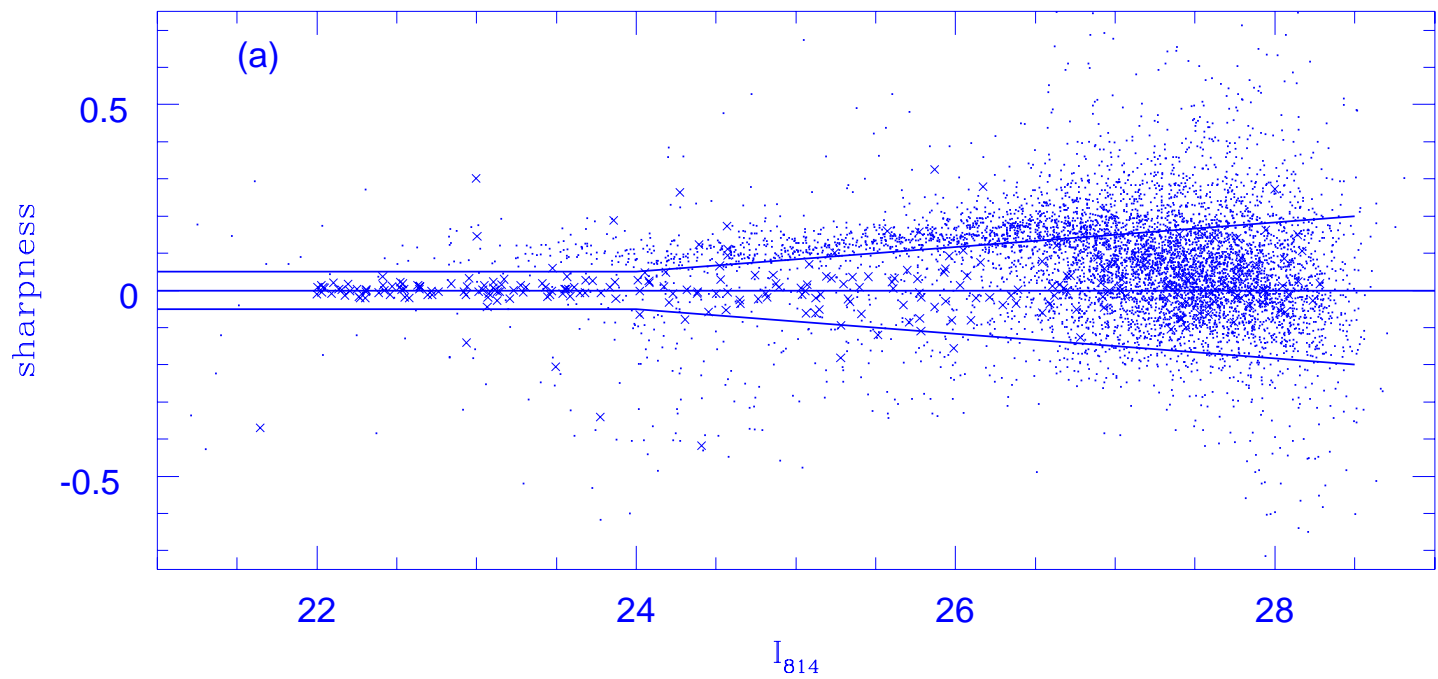
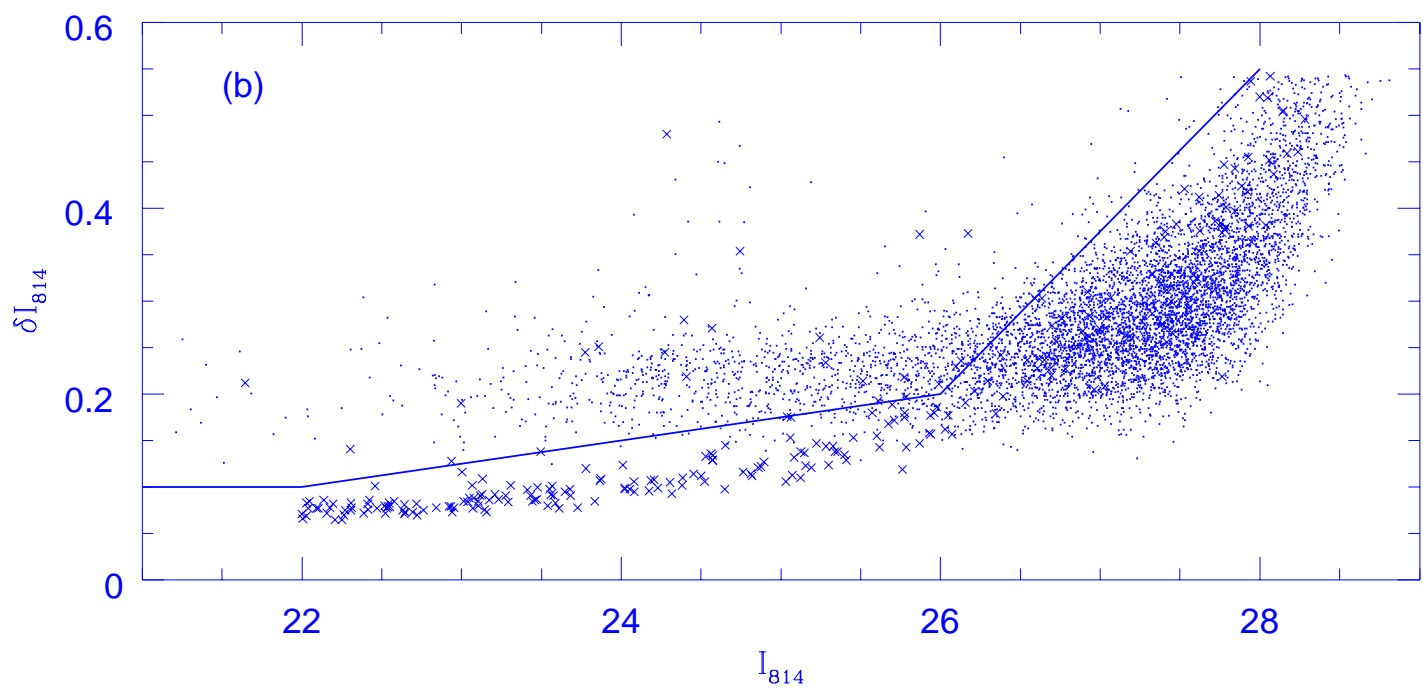
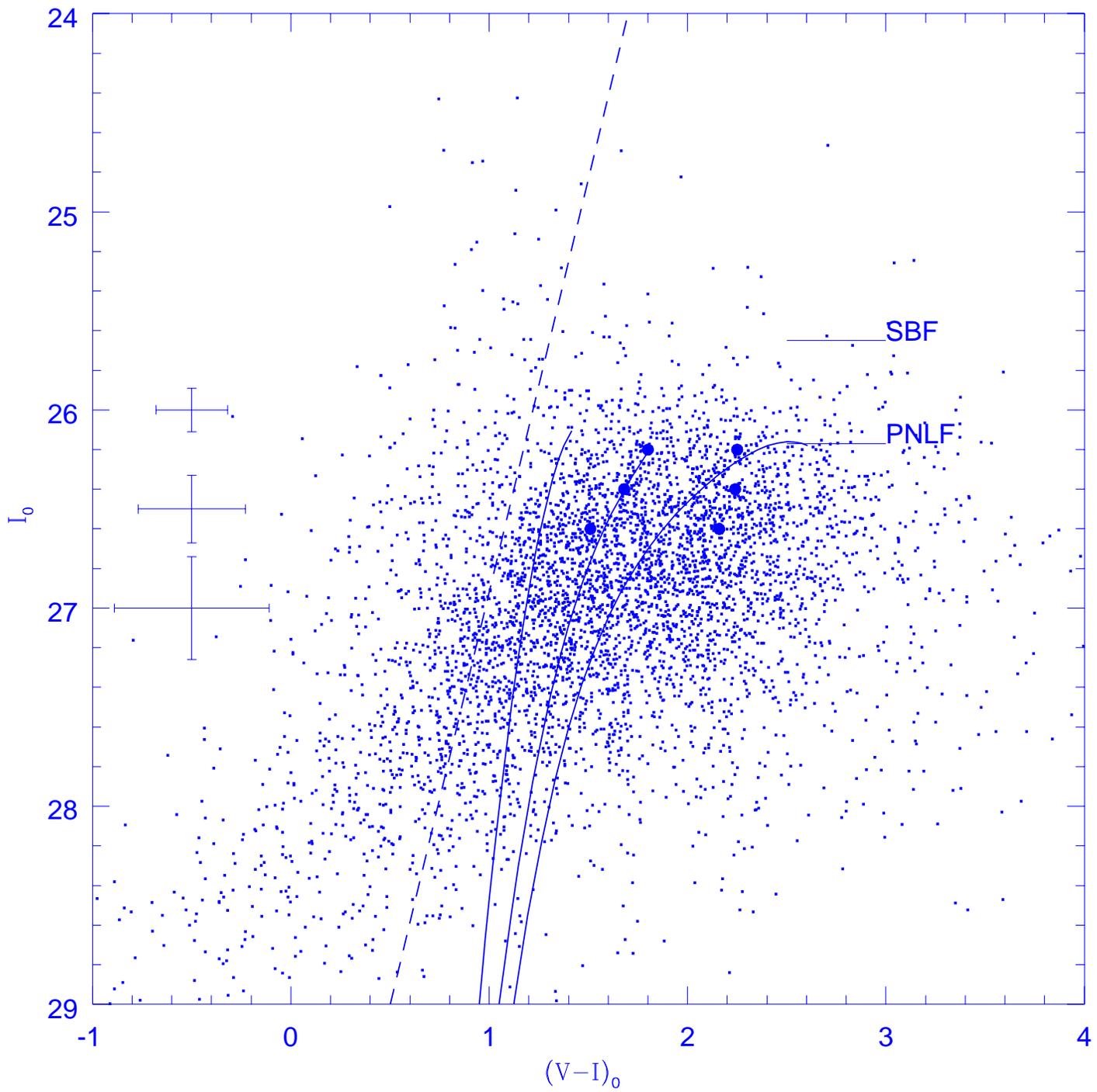
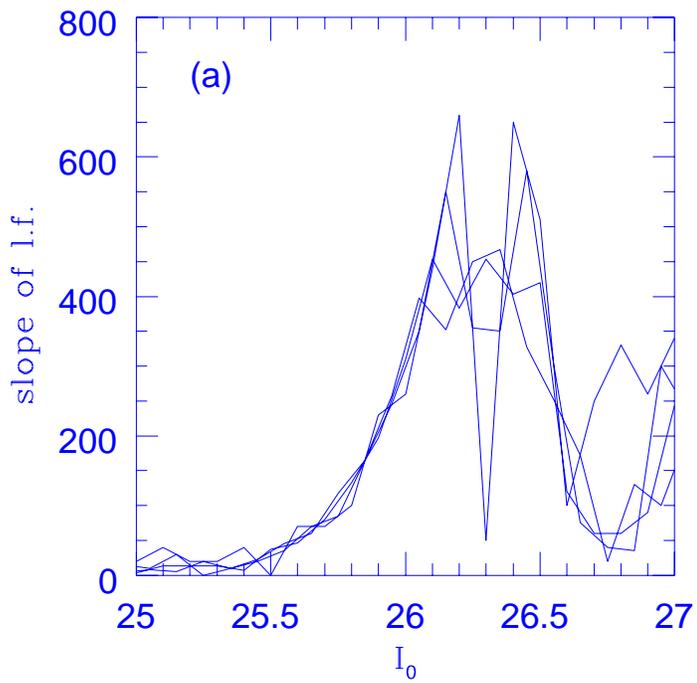
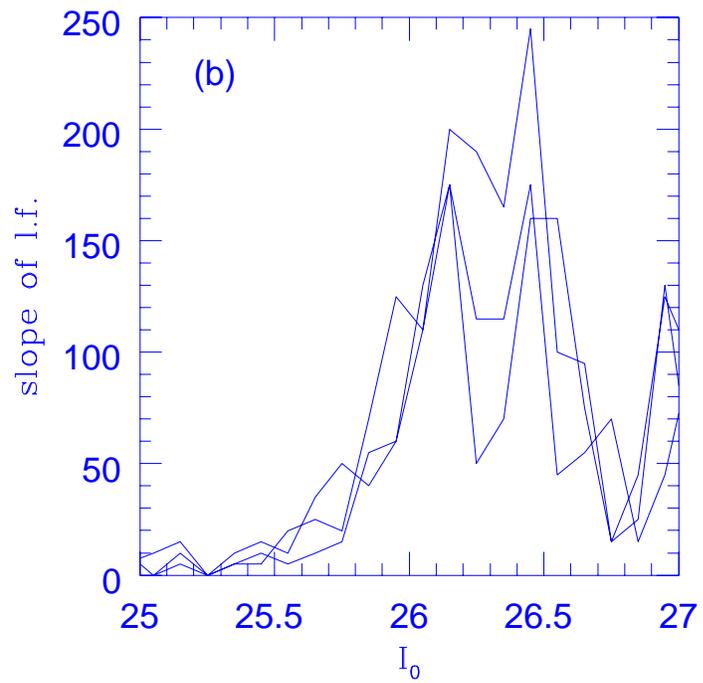
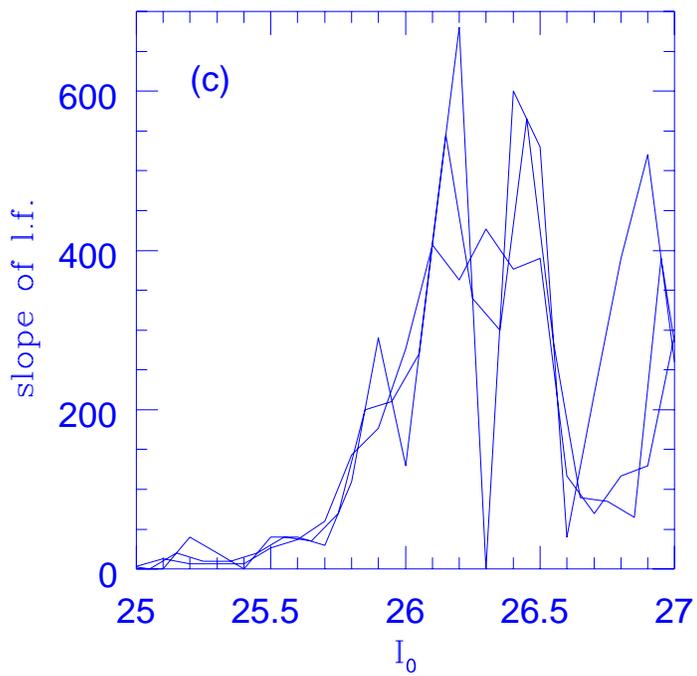
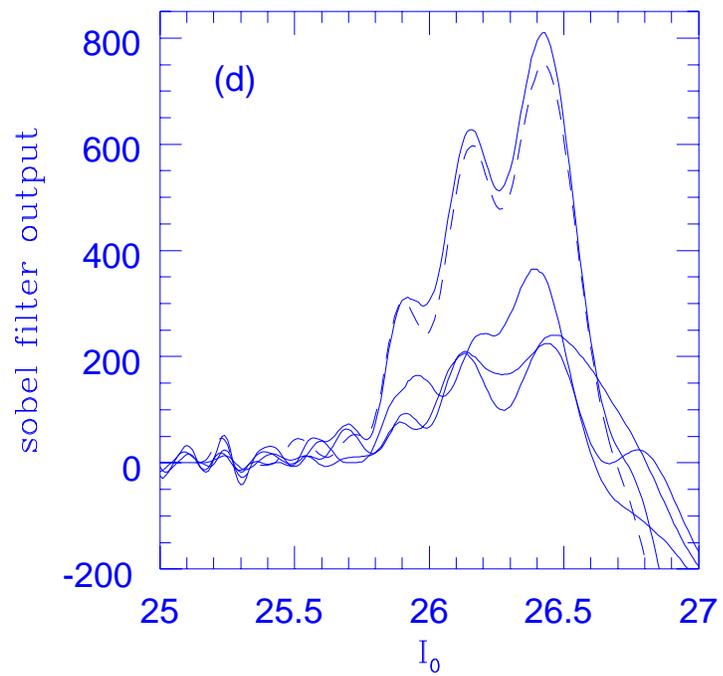
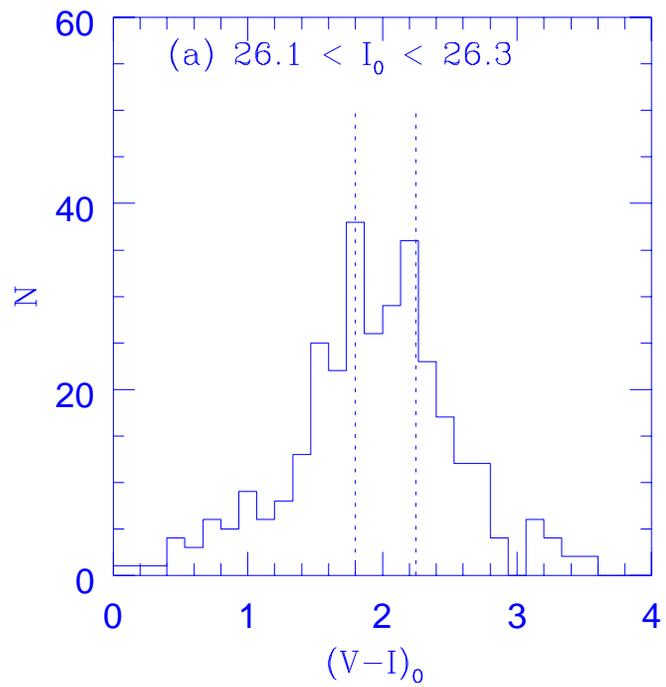
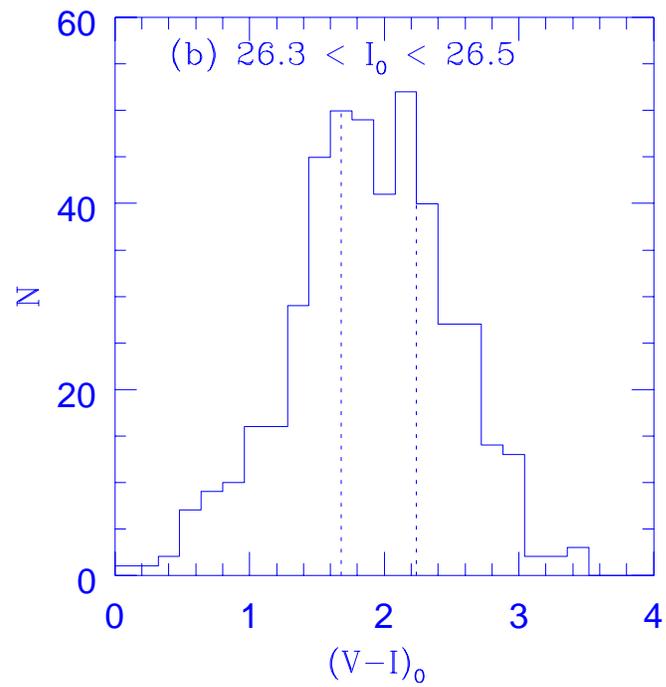
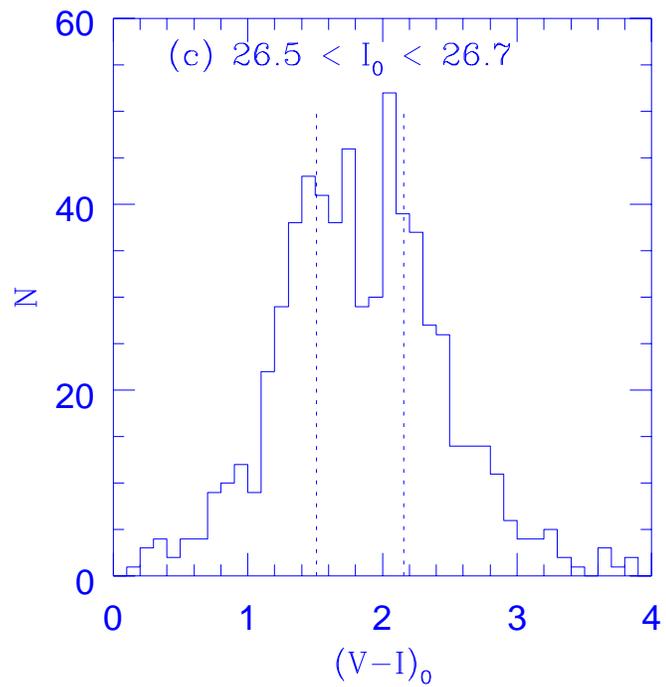
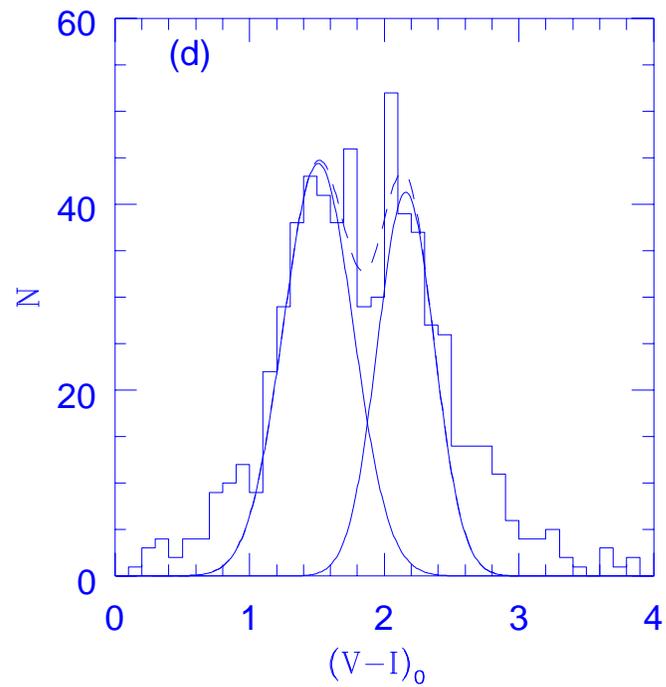
\end